\documentclass[12pt, journal, twocolumn]{IEEEtran}
\usepackage{amssymb,amsmath}
\usepackage{cite}
\usepackage{graphicx,subfigure}
\usepackage{psfrag}
\usepackage{url}
\usepackage{booktabs}
\usepackage[latin1]{inputenc}
\usepackage[absolute,overlay]{textpos}
\usepackage[latin1]{inputenc}
\usepackage{color}

\usepackage{array}
\newcolumntype{L}[1]{>{\raggedright\let\newline\\\arraybackslash\hspace{0pt}}m{#1}}
\newcolumntype{R}[1]{>{\raggedleft\let\newline\\\arraybackslash\hspace{0pt}}m{#1}}
\usepackage{arydshln}

\begin{document}	
	\title{Ultra-low Latency, Low Energy and Massiveness in the 6G Era via Efficient CSIT-limited Schemes}
	\author{\IEEEauthorblockN{Onel L. A. L\'opez,
			Nurul H. Mahmood,
			Hirley Alves,
			Carlos M. Lima, 
			Matti Latva-aho
		}
		\thanks{Authors are all with the Centre for Wireless Communications (CWC), University of Oulu, Finland. \{onel.alcarazlopez, carlos.lima, nurulhuda.mahmood, hirley.alves, matti.latva-aho\}@oulu.fi.}
		\thanks{This work is partially supported by Academy of Finland 6Genesis Flagship (Grants n.318927, n.319008, n.307492).}
	    \thanks{\copyright 2020 IEEE. This paper has been accepted for publication in IEEE Communications Magazine. Personal use of this material is permitted. Permission from IEEE must be obtained for all other uses, in any current or future media, including reprinting/republishing this material for advertising or promotional purposes, creating new collective works, for resale or redistribution to servers or lists, or reuse of any copyrighted component of this work in other works.}	
}						
	\maketitle
	\begin{abstract}
		Channel state information (CSI) has been a key component in traditional wireless communication systems. This might no longer hold in future networks supporting services with stringent quality of service constraints such as extremely low-latency, low-energy and/or large number of simultaneously connected devices, where acquiring CSI would become extremely costly or even impractical. 
		We overview the main limitations of CSI at transmitter side (CSIT)-based designs toward the 6G era, assess how to design and use efficient CSIT-limited schemes that allow meeting the new and stringent requirements, and highlight some key research directions. We delve into the ideas of efficiently allocating pilot sequences, relying on statistical CSIT and/or using location-based strategies, and demonstrate viability via a selected use case.
	\end{abstract}
%	\vspace{-6mm}
	\section{Introduction}\label{Int}
	%
%	\vspace{-1mm}
	Both industry and academia have their sights set on a data-driven sustainable society, enabled by near-instant unlimited green connectivity toward the 6G era. The main challenges ahead are due to the need to support a wide range of heterogeneous characteristics as surveyed in \cite{Mahmood.2020}, and conflicting   requirements, \textit{e.g.}, ultra-low latency, ultra-reliability, low-energy, massive connectivity support, ultra-high data speed. These challenging requirements demand a complete rethinking of the usual system optimization approaches and design criteria \cite{Mahmood.2020,Popovski.2018}. For instance, accurate channel state information (CSI), which is acquired via proper pilot training  transmissions, is a key component for modern communication systems. However, while CSI acquisition costs are usually negligible in traditional broadband-orientated services, this may no longer hold under strict constraints on latency, energy expenditure, and/or when serving a huge number of devices. 
	
	CSI at the receiver side (CSIR) allows using coherent receiving schemes, which optimize the data decoding procedures. The associated costs are $i$) extra communication delay since data decoding does not start until CSIR acquisition has been completed, and $ii$) extra energy consumption to carry out not only the estimation phase but also the coherent decoding. The situation becomes more challenging under grant-free access schemes, in which there is no reservation phase, hence there might be collisions among users' transmissions especially in dense deployments. Compressed sensing techniques emerge such that users prepend sequences to the transmitted data, which then are used for both activity detection and CSIR acquisition. CSIR acquisition problem is even more critical in time-varying channels, \textit{e.g.}, under high mobility conditions, for which non-coherent schemes may offer better performance.
	
    The challenges and associated costs are more stringent when CSI is required at the transmitter side (CSIT),  \textit{e.g.}, for optimal multi-user spatial precoding and/or intelligent resource allocation schemes. In fact, CSIT acquisition is severely limited toward realizing ultra-reliable low-latency communication (URLLC) \cite{Popovski.2018}, 
	and serving low-power massive deployments \cite{Lopez.2019_Alves}, thus, fostering efficient CSIT-limited schemes \cite{Yin.2013,Lopez.2019_Alves,Lopez.2020,Qiu.2018}. It is worth mentioning that decades ago, before introducing adaptive modulation and coding schemes and multi-antenna beamforming, wireless communication systems operated without CSIT. However, the stringent performance requirements of current and future systems make the design and adoption of efficient CSIT-limited schemes more challenging than ever. 	
	In this paper, we focus  on this, while our four-fold contributions are as follows: \textit{i}) we discuss the costs and challenges of the CSIT acquisition procedures toward meeting the 6G era requirements (Section~\ref{sec:csi_cost}); \textit{ii}) we overview some solutions under consideration, and propose novel ideas to avoid the need of instantaneous CSIT, hence enabling ultra-low latency, low-energy and massive communications (Section~\ref{sec:csi_free_sols}); \textit{iii}) we provide a study case on how to power efficiently a large number of energy harvesting (EH) devices via CSIT-free schemes (Section~\ref{C}); and \textit{iv}) we discuss open problems and future research directions toward efficient CSIT-limited schemes in Table~\ref{table}.		
    %
%	\vspace{-3mm}
	\section{On the CSIT acquisition costs} \label{sec:csi_cost}
	%
%	\vspace{-1mm}
	Typically, CSIT-based solutions come with the strong assumption that channel information is readily and widely accessible in the network. However, acquiring such CSIT is costly and even infeasible in most practical deployment scenarios. In this section, we overview the associated costs and challenges. For simplicity, let us denote by $S$ the node with a data message ready for transmission, while $D$ is the intended receiver. 
	%	
%	\vspace{-4mm}
	\subsection{Latency cost}\label{latency} 
	%
%	\vspace{-1mm}
	In frequency-division duplex (FDD) systems, where downlink and uplink channels operate on different frequencies, $S$ transmits pilot signals while $D$ uses them to estimate the channel. Then, $D$ sends the information back to $S$, thus, introducing an additional delay and error source \cite{Popovski.2018,Qiu.2018}. If CSIT acquisition time window is limited, the accuracy (directly related to 	energy and orthogonality level of the pilot signal, along with the allowed quantization error) and/or reliability (determined by the error performance of the feedback link) can be seriously compromised. 
	When $D$ is a very simple device, \textit{e.g.}, an Internet of Things (IoT) sensor, channel estimation may be unaffordable due to its limited (if any) baseband signal processing capabilities. Training overhead grows with the number of transmit antennas, thus,  it becomes troublesome when $S$ is equipped with large antenna arrays \cite{Qiu.2018,Adhikary.2013}.
	
	On the other hand, in time-division duplex (TDD), where uplink and downlink channels share the same frequency resources, the above issues can be addressed by exploiting channel reciprocity. However, channel reciprocity is sensitive to hardware impairments especially when $S$ and $D$ are very different devices. In general, calibration methods are required to mitigate the channel mismatches. Although they may run over a large time-scale, \textit{e.g.}, each 10 mins in centralized multi-user multiple-input multiple-output (MIMO) systems sharing a common clock, they could be required on a frame-basis in distributed architectures \cite{Rogalin.2014}. Receiver mobility, which leads to time-varying channels, especially between downlink and uplink, makes channel tracking rather difficult and costly, and constitutes another important challenge.
	
	Time requirements for CSIT acquisition are~not trivial, not even in TDD, and they grow proportionally with the pilot sequences' length, which is usually set to avoid non-orthogonality in multi-user uncoordinated setups (see Section~\ref{sheduling}). 
	%
%	\vspace{-4mm}
	\subsection{Energy cost} \label{energy}
	%
%	\vspace{-1mm}
	% 
	In general, CSIT estimation accuracy  depends on the   energy expended in associated procedures. The energy expended by an IoT device $D$ is proportional to the pilots energy in TDD, which is explicitly determined by the training time and power, \textit{i.e.}, \textit{energy $=$ time $\times$ power}; while the energy consumed by $D$ in FDD comprises that used during the channel estimation (proportional to the training time) and feedback transmission (equivalent to the training power in TDD). Hence, energy expenditure and CSIT accuracy, which affect the subsequent data transmission phase, can be traded off. The problem of optimizing the system energy efficiency measured as the ratio (or some other function) between \textit{useful data }throughput (or rate) and power (or energy) subject to imperfect CSIT has been frequently considered in the literature.
	However, in low-power IoT, low energy operation is frequently much more important than data rate or throughput metrics. The main reasons are that \textit{i}) transmissions are mostly sporadic, \textit{e.g.}, due to event-driven traffic; \textit{ii}) transmission rate is often intrinsically small and fixed; and \textit{iii}) such simple devices are battery-driven and or rely on weak energy sources, \textit{e.g.}, as in wireless-powered communication networks. The main issue lies on optimizing the network life-time avoiding frequent channel training procedures that drain significant energy resources. 
	%
%	\vspace{-5mm}
	\subsection{Scalability}\label{sheduling}
	%
%	\vspace{-1mm}
	Training overhead may remain fixed in coordinated systems but at expense of costly scheduling (unless the traffic is periodic or semi-periodic) and synchronization. Meanwhile, in non-coordinated single-cell/centralized networks, the overhead grows
	\begin{itemize}
		\item  (at least) logarithmically with the network size in TDD to guarantee pilot signals' orthogonality, hence, avoiding collisions; or
		\item linearly with the number of antennas in $S$ in FDD to estimate the channel from each of the transmit antennas.
	\end{itemize}
	Meanwhile, the situation is more critical in multi-cell networks and distributed antenna systems (DAS) since the training overhead grows also with the system size/density. In addition, advanced estimators and intelligent pilot scheduling may be needed to acquire accurate channel estimates and enable,  \textit{e.g.}, efficient massive MIMO \cite{Sanguinetti.2020}. 
	%
%	\vspace{-4mm}
	\section{CSIT-limited solutions for $6$G systems}\label{sec:csi_free_sols}
	% 
%	\vspace{-1mm}
	5G systems have already introduced services with ultra-low latency, energy efficiency  and/or massive support requirements for which the  aforementioned CSIT acquisition costs (see Section \ref{sec:csi_cost}) constitute a difficult obstacle to overcome. However, the challenges are undoubtedly  strenuous toward the 6G era since performance requirements will be more heterogeneous, conflicting and stringent than ever, \textit{e.g.}, 3D wide coverage of \textit{zero-energy} (ZE) devices, extreme low-latency ultra-reliable broadband services \cite{Mahmood.2020}. This will be  motivated by new use cases, such as Internet of Senses and ZE IoT, and the wide adoption of challenging technologies, for example, intelligent reflecting surfaces (IRS), EH, backscatter communications, distributed ledger \cite{Mahmood.2020}.
    Next, we discuss promising strategies to enable ultra-low latency, low-energy and/or massive extreme support in 6G systems. It is worth mentioning that expected advances  on  ultra-low power circuit designs,  machine learning (ML)/artificial intelligence (AI) algorithms and positioning techniques with centimeter/ millimeter-level accuracy  will  facilitate the adoption of such enabling strategies toward the 6G era. 
	%	
%	\vspace{-4mm}
	\subsection{Efficient pilot sequences allocation}\label{pilots}
	%
%	\vspace{-1mm}
	The more network pilots used for CSIT acquisition, the more stringent the time/energy demands, thus, assigning orthogonal pilots becomes inefficient/unaffordable as the network grows. By tightly constraining the pilot pools, not only the required time/energy expenditure may be minimized but alternatively the network is allowed to scale up. The use of non-orthogonal pilot sequences may be viable, by resorting to, \textit{e.g.}, grant-free access schemes \cite{Carvalho.2017}, which in turn increase the collision probability and consequently cause a longer effective delay. Thus, the probability of such collision events must be minimized. For instance, the authors in \cite{Adhikary.2013, Yin.2013} propose to cluster the destinations $\{D\}$ based on their channel covariance matrices such that the pilots are reused among the users having sufficiently orthogonal channel subspaces, thus, mitigating the pilot contamination effect in the massive MIMO regime. Clustering approaches can also be effectively employed in the non-massive antenna regime, \textit{e.g.}, devices with similar traffic profiles can be clustered together and share a specific pool of pilots (which may be accessed randomly as shown in \cite{Carvalho.2017}). In one implementation, devices with small activation probabilities can be grouped in clusters with smaller pilot pools, when compared to clusters enclosing more frequently \textit{active} devices which conversely need bigger pilot pools. In this way, the chances of collision  decrease. Such strategy, enabled by novel risk-sensitive ML/AI algorithms, could be further enhanced by considering the features and requirements of the different services to be supported. For instance, more pilots should be assigned to clusters with greater chances of requiring URLLC services, which are more sensitive 	to collisions, while ZE devices should be favored to some extent as well since their transmissions may be intrinsically limited in number/energy.
	\begin{figure}[t!]
		\centering  
		\qquad\includegraphics[width=0.95\columnwidth]{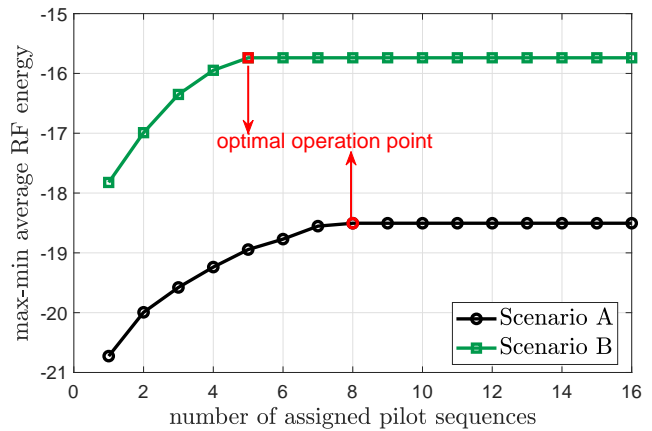}
		\vspace{-3mm}
		\caption{Max-min average RF energy available at end-devices. We consider a set of 16 EH devices and one PB equipped with $4$ antennas. Channel experiences quasi-static Rician fading with line-of-sight (LoS) factor of $3$ dB. each device $D_i,\ i=1,\cdots,16$ experiences a path-loss of $50+i/2$ dB (Scenario A) or $44+i$ dB (Scenario B). Transmit power is normalized.}	
		\label{Fig1}
		\vspace{-3mm}
	\end{figure}
	
	In some cases, it could be even preferable to not assign pilots to certain users at all. One example scenario is the fair wireless powering of EH devices. Consider a multi-antenna power beacon (PB) $S$, which is sending radio frequency (RF) signals for powering a set $\{D\}$ of low-power EH nodes, \textit{i.e.}, a wireless energy transfer (WET) setup. $S$ uses beamforming based on instantaneous CSIT to improve the statistics of the devices' harvested energy. Specifically, for maximum fairness the beamforming problem can be cast to maximize the minimum energy that can be harvested by any device. The traditional approach would rely on the channel information of the entire set of devices, which could be large, hence hindering an energy-efficient pilot scheduling. However, the wireless powering efficiency decays quickly with the distance, following a power-law. Therefore, farthest users' channels are expected to dominate the beamforming design and channel training can be reduced accordingly, which is illustrated in Fig.~\ref{Fig1}. Also, as the LoS increases, a smaller number of pilot sequences are necessary since the channels become more deterministic.
    %
%	\vspace{-4mm}
	\subsection{Exploiting the statistics/structure of the channel}\label{structure}
	%
%	\vspace{-1mm}
    %
	Not only the multipath structure or statistical knowledge of the transmit channels may be used to efficiently assign the pilots \cite{Adhikary.2013, Yin.2013}, but  the spatial beamforming can entirely rely on these features \cite{Qiu.2018}. Statistical CSIT remains accurate  over a  large time scale in static setups and can be even predicted based on ML/AI and other statistical methods. This allows performing CSIT acquisition procedures when delay and/or energy constraints are favorable, while relying on the learned channel statistics when constraints become stricter and CSIT auxiliary procedures are unaffordable. In this way, the number of pilots may decrease, thus additionally achieving the benefits discussed in Section~\ref{pilots}. 	
		
	Statistical beamformers have attracted much attention from the early GSM systems up to date in massive MIMO \cite{Qiu.2018},  although they have not succeeded completely. Probably because their performance is strictly sub-optimal for broadband-orientated services, where the impact of instantaneous CSIT acquisition overhead is usually negligible,
	and/or when channels tend to fade significantly. However, the former does not hold already for many services and much less toward the 6G era as discussed in previous sections, while the latter may not be an issue as the network densifies.	
	
    Let us consider a simple setup in which $S$ is equipped with four antennas and serves four single-antenna users in the downlink. $S$ uses a precoding vector such that the minimal average signal-to-interference-plus-noise ratios (SINR)	is maximized. In case the nodes are under outage-latency constraints as in URLLC, one can map the constraints to the corresponding average SINR requirements since channel statistics are known. 
	\begin{figure}[t!]
		\centering  
		\qquad\includegraphics[width=0.95\columnwidth]{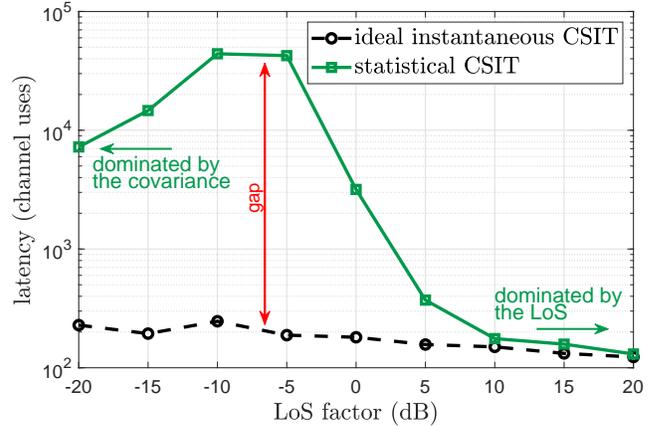}
		\vspace{-3mm}
		\caption{Latency vs LoS factor for a target error probability of $10^{-4}$. We consider a quasi-static Rician fading with per-link average signal-to-noise ratio of $6$ dB. $S$ is equipped with an uniform linear array (ULA) whose antenna elements are separated by half-wavelength, while the azimuth angle relative to the boresight of the transmitting antenna array with respect to end-devices $\{D_i\}$ is $0^\circ,30^\circ,60^\circ$ and $90^\circ$. We assume real covariance matrices with exponential correlation with parameter $i/5$ for $D_i$.}		
		\label{Fig2}
		\vspace{-6mm}
	\end{figure}
	In Fig.~\ref{Fig2}, we can observe that the performance of the statistical precoder improves as the LoS (direct) component becomes stronger in comparison to the other (scattered) multi paths. 
	This is extremely relevant in dense  deployments and DAS wherein communicating pairs experience less path-loss attenuation (due to shorter separation distances) which, as a result, increases the corresponding received power through the LoS component (direct link). Thus, an appropriate statistical precoder may reach near-optimum performance in such propagation conditions. Meanwhile, future wireless systems will greatly benefit from massive deployments of IRS. These metasurfaces, which consist of numerous passive reflecting elements,  are deployed to assist blocked $S\rightarrow D$ communication channels by providing an alternative software-controlled LoS path. Since IRS are mostly passive by nature, the design of their interaction matrices cannot rely on frequent channel training (\textit{e.g.}, see \cite{Taha.2020} for a problem  overview and a novel solution). Actually,  as suggested by Fig.~\ref{Fig2} results, training can be considerably reduced by just relying on LoS information, which is mainly influenced by the channel average and the network geometry.

	Finally, statistical CSIT and readily available side-information, \textit{e.g.}, battery charge information, can be intelligently exploited for efficient resource allocation, even in single antenna setups, under probabilistic quality-of-service (QoS) constraints.
	%
%	\vspace{-4mm}
	\subsection{Location-based strategies}\label{locat}
%	\vspace{-1mm}
	When CSIT acquisition via  pilot transmissions becomes costly or even impractical, location information may be used to effectively reduce signaling overhead and feedback delays. Location information is valuable, \textit{e.g.}, as an indication of the received power and experienced interference levels,  as a node association criterion, and to enable efficient relay selection and routing strategies, and specially in future IoT use cases wherein heterogeneous sources of information dynamically interact in a distributed manner \cite{Taranto.2014}. In DAS and multi-connectivity scenarios, location information can be used to estimate the signals' propagation delays coming from different transmit nodes $S$, and adapt the transmit signals' timing accordingly. Location-based solutions achieve lower spectral efficiency in highly dynamic scenarios, though at a lesser cost in terms of measurements and signaling exchange (thus, more time/energy-friendly), when compared to  typical CSIT-based approaches~\cite{Botsov.2015}.
	
	There also exist solutions that directly connect the location information to the radio channel, \textit{e.g.},  cell-center/edge users share resources in low-power wide-area network deployments based on their position. Location-based multicasting is possible therein, which increases packet delivery reliability by reducing contention.
	Location-domain channel representation based on the geometric constraints from the local scattering \cite{Molev.2019} is used to effectively estimate the channel with a reduced number of pilots in distributed MIMO systems. 
	
	Location information has been recently used in ML/AI frameworks to infer the channel quality even in locations where no measurement information was previously available. In that regard, a Gaussian process approach is used together with received power measurements and temporal correlation database so as to predict location-dependent channel quality and support resource allocation strategies \cite{Taranto.2014}. In general, location-based CSIT estimation is coarse and statistical by nature since small-scale fading is extremely difficult (if even possible) to predict. However, ML/AI-enabled methods can take advantage of much more diverse data to accurately predict higher order CSIT statistics toward 6G systems, thus; reducing prediction uncertainty and allowing more efficient  decision-making/resource allocation. We envision application scenarios as the one illustrated in Fig.~\ref{Fig3}, where the base station $S$ needs to efficiently serve two devices $D_1$, $D_2$ whereas pilot-assisted CSIT acquisition procedures are not accessible. ML/AI methods can be fed with data known to influence the RF propagation conditions such as: 
	\begin{figure}[t!]
		\centering  
		\qquad\includegraphics[width=0.95\columnwidth]{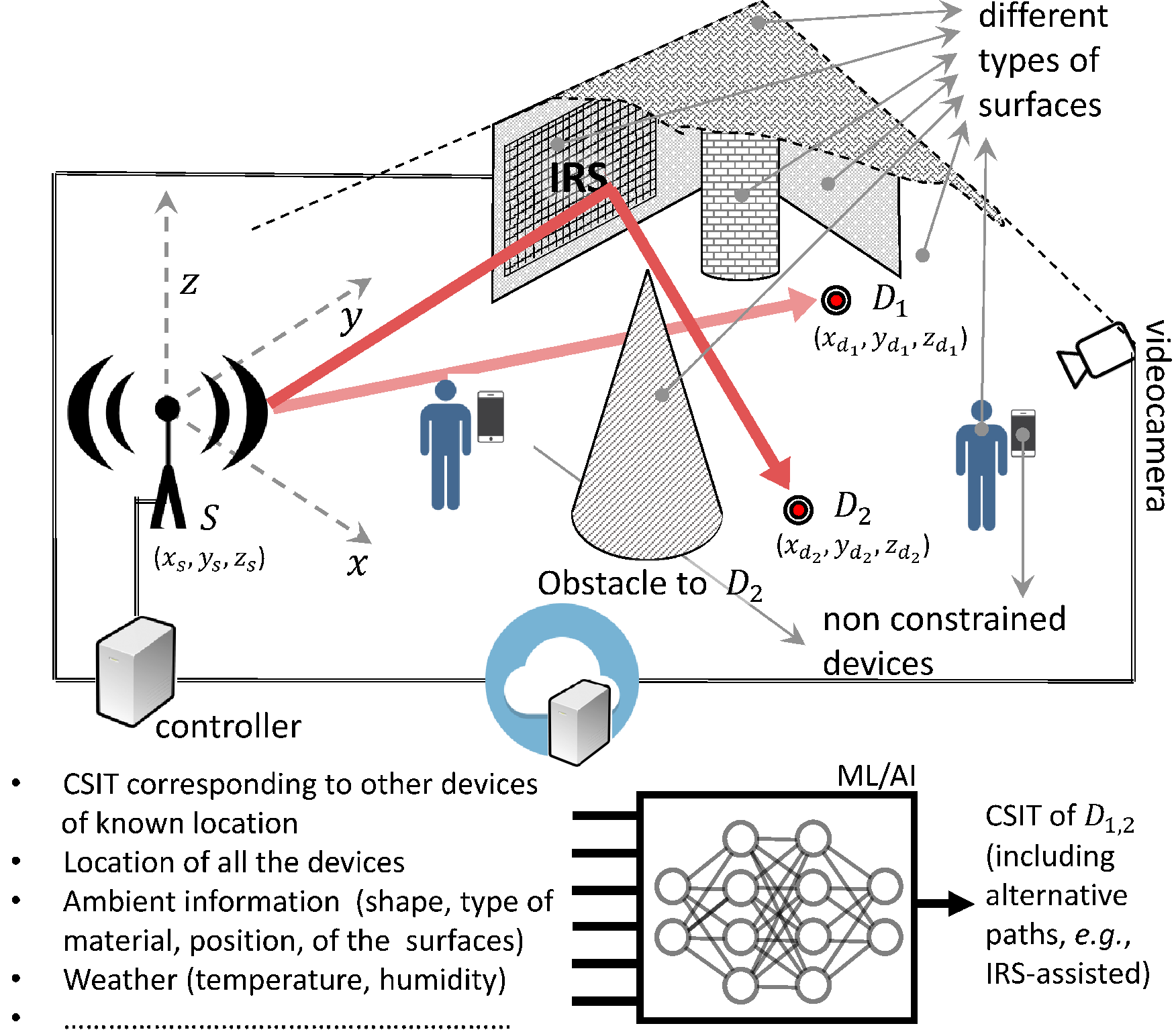}
		\vspace{-2mm}
		\caption{Representative scenario for ML/AI-enabled CSIT-acquisition. Location information is crucial input data. The ML/AI block may be at the controller or in a cloud server according to the application.}	
		\label{Fig3}
		\vspace{-4mm}
	\end{figure}
	\begin{itemize}
	    \item geometry and construction materials of the environment (which influence mostly the reflections, diffraction, penetration, scattering phenomena, and can be obtained via imaging, available locally or from the cloud);
	    \item weather conditions (humidity and temperature vary the materials' features, and can be obtained  from cloud or locally based on sensing);
	    \item in addition to actual (updated) data mapping, the channel information at certain locations (which can be obtained by $S$ from other non-constrained devices). Note that this data is used as input to the ML/AI block only  when  updated, otherwise it is used for training.
	\end{itemize}
	If channels are predicted blocked, $S$ can take advantage of alternative paths enabled by IRS deployments. These alternative channels may be estimated, thus enabling efficient resource allocation at $S$. 
	\vspace{-3mm}
	\section{Case Study: CSIT-free massive WET} \label{C}
	\vspace{-1mm}
	For the most stringent scenarios, such as massive WET for ZE devices,  CSIT-free schemes may be extremely beneficial. Note that relying on instantaneous CSIT with reduced number of pilots as illustrated in Fig.~\ref{Fig1}, is not completely convenient here since the farthest (more energy-limited)  users would still required to spend valuable energy resources for channel training. Under strict latency constraints, statistical beamforming is viable, similarly to our previous discussions around Fig.~\ref{Fig2}, however, the energy expenditure in the auxiliary procedures is not completely avoided. Finally, the denser the network, the more efficient the CSIT-free schemes. 
	
	Two promising  CSIT-free  schemes for massive WET are \cite{Lopez.2019_Alves,Lopez.2020}: \textit{i}) all antennas at once ($\mathrm{AA}$), under which the PB transmits the same signal simultaneously with all antennas; and \textit{ii}) switching antennas ($\mathrm{SA}$), under which the PB transmits by one antenna at a time such that all antennas are used during a coherence block, thus just one RF chain is required, reducing circuit power consumption, hardware complexity and the operational cost. Their origins date back to the 90s but in a wireless communication context with limited multi-user support \cite{Mah.1999}. However, more recent advances on MIMO and its geometric modeling, signal processing and software-controlled rotors, may fully realize their potential in a massive WET context. 

\begin{figure}[t!]
		\centering  
	\ 	\includegraphics[width=0.94\columnwidth]{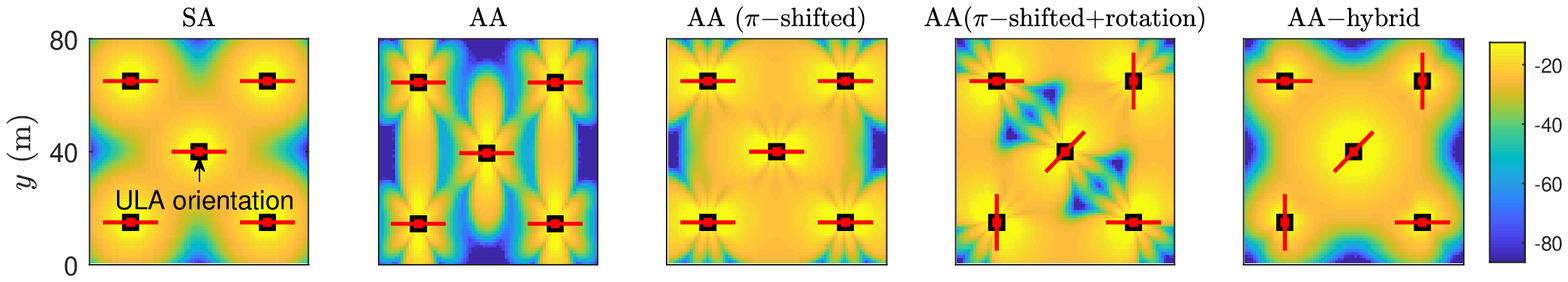}\\
		\vspace{2mm}
		\ \   \includegraphics[width=0.94\columnwidth]{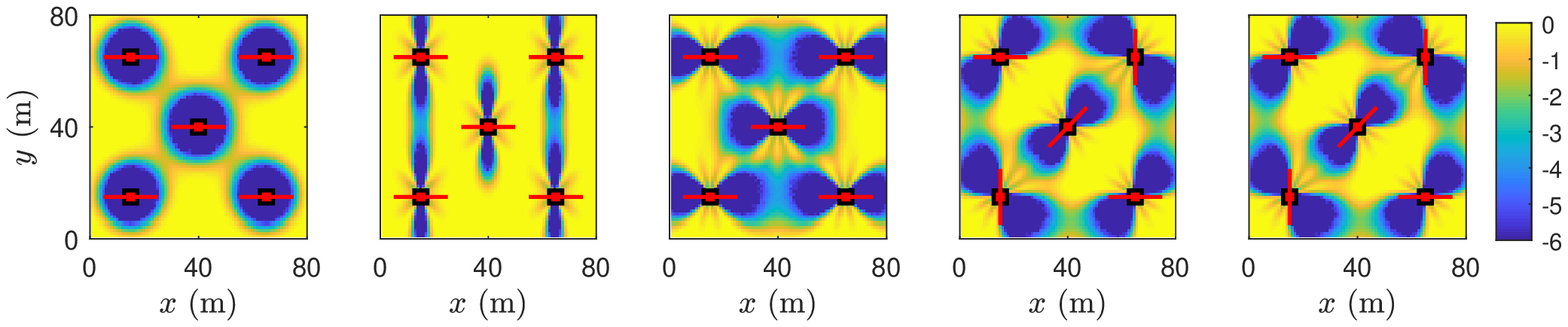}
		\vspace{-2mm}
		\caption{Heatmap of $(a)$ average harvested energy in dBm (first row) and $(b)$ $\log_{10}$ of the energy outage probability (second row) in a $80\times 80$ $\mathrm{m}^2$  area. Five PBs equipped with 4 ULA antennas transmit independent signals. We consider a log-distance path-loss model with exponent $3$, while the non-distance dependent losses are $26$ dB. Channels are under quasi-static Rician fading with LoS factor $10$ dB. EH circuitry operates with sensitivity, saturation and conversion efficiency of $-22$ dBm, $-8$ dBm and $35\%$, respectively. 
		}		
		\vspace{-6mm}
		\label{Fig4}
	\end{figure}

	$\mathrm{AA}$ has been proven to benefit greatly from spatial correlation and LoS; while $\mathrm{SA}$ does not improve the average statistics of the available RF power but it does provides full diversity gain, making it especially desirable under non-LoS conditions. For the first time, herein we evaluate previous CSIT-free schemes in a multi-PB scenario. Results in the form of heatmap for the average harvested energy and energy outage probability are shown in Fig.~\ref{Fig4}. $\mathrm{SA}$ provides a uniform performance along the area in terms of both average harvested energy and energy outage probability, while the performance of $\mathrm{AA}$ depends strictly on the ULA orientation. By shifting the signals phases by $\pi$ in consecutive antenna elements (as recommended by \cite{Lopez.2020}) and/or intelligently rotating the antennas, $\mathrm{AA}$'s performance can be improved. While the original $\mathrm{AA}$ allows the EH devices to harvest  energy not less than $-20$ dBm on average in $15\%$ of the area, such coverage can be increased to $23\%$, if $\mathrm{SA}$ is used, or even to $30\%$, if the two more  advanced $\mathrm{AA}$ schemes are utilized. 
	\vspace{-3mm}
	\section{Conclusions \& Outlook }
	\vspace{-1mm}
	We discussed the enormous costs of acquiring instantaneous CSIT toward future communication systems, where stringent QoS constraints become extremely challenging. The costs were analyzed in terms of latency and energy expenditure, while the scalability with the number of devices was also considered. We discussed some feasible approaches that range from efficient allocations of the pilot sequences up to schemes relying on statistical CSIT or location-based ML/AI-enabled predicted CSIT. We showed that \textit{i}) in low-power massive setups, not all the EH devices  may need to be served with CSIT, hence saving valuable energy resource; while \textit{ii}) in a downlink communication setup, statistical CSIT schemes approach the optimum performance as the LoS factor increases, making them favorable as the network densifies. We  highlighted the benefits from efficiently-designed CSIT-free WET schemes in multi-PB setups. Results revealed performance improvements are reachable via properly shifting the signals' phases transmitted by each antenna and/or rotating the antenna array. Key challenges and research directions are identified in Table~\ref{table}.
	\begin{table*}[!t]
	\centering
	\caption{Some Research Directions}
	\vspace{-3mm}
	\begin{tabular}{
			 L{0.266\textwidth} L{0.415\textwidth} L{0.106\textwidth}
			 L{0.125\textwidth}}
		\toprule \\[-4mm]
		\textbf{Challenge} & \textbf{Candidate Approaches} & \textbf{Requirements} & \textbf{Tools}
    	\\[-0.8mm] 
		\midrule
		\textit{Mitigate the pilot shortage phenomena:}
		In massive IoT deployments, appropriate allocation strategies are required to deal with the finitude of the pilot pools.
		& Intelligent and dynamic network clustering to assign different  pilot pools according to the identified traffic features and QoS requirements. Soft-clustering, where there is only a probabilistic certainty of belonging to certain cluster, may be needed. Also, intelligent decision-making based on risk mitigation.
	    &  traffic, QoS, location information, and/or statistical CSIT  &  ML/AI clustering methods, hidden Markov models, risk analysis \\
    	\hdashline
    	\textit{Statistical beamforming in DAS:} Geographically-distributed antennas improve the network spectral efficiency by mitigating the  large-scale fading.  Novel beamforming designs exploiting statistical channel information are required for DAS. & DAS can be implemented in several ways, such as multi-connectivity transmissions and the novel cost-efficient radio stripe systems. The candidate beamforming solutions must consider the characteristics/challenges of the specific DAS deployment. In many cases, standalone statistical beamforming designs could be adapted to DAS as centralized algorithms, but the use of distributed solutions is usually preferable.  & local and/or global statistical CSIT, QoS requirements & (statistical  /probabilistic) optimization, game theory, meta distribution, network calculus, distributed algorithms \\
    	\hdashline
     	\textit{Beamforming under heterogeneous CSIT availability:} Due to network heterogeneity, $S$ may need to simultaneously serve devices for which \textit{i})  CSIT acquisition procedures are affordable, \textit{ii}) there is statistical knowledge of their channels but training is not currently possible, and \textit{iii}) CSIT is unavailable. & Under heterogeneous CSIT availability, hybrid beamformers are necessary, while they need to account also for the different (possibly of different nature also)  QoS requirements. Probabilistic optimization approaches are necessary, as well as novel networking mechanisms to incentive information sharing among network devices. & local/global full/limited CSIT, QoS requirements, location, traffic information & (statistical /probabilistic) optimization, distributed ledger technology, risk analysis, cooperation\\
	    \hdashline
    	\textit{Beamforming for  highly mobile networks:} In highly mobile networks, acquiring any kind of CSIT may be completely unfeasible since the the channel coherence time shrinks considerably. & Location-based beamforming allows directing the energy to the desired LoS direction by exploiting proper phase-shifting between the transmit antenna array elements. However, an accurate prediction of the receiver's LoS direction is required. An efficient design must deal with mobility-tracking prediction error, account for the impairments introduced by the Doppler effect, and address the circumstances in which the LoS link is blocked and strong non-LoS links can be exploited. & environment geometry, location, QoS requirements & ML/AI-based location estimation/prediction, ray tracing, cooperation \\
    	\hdashline \textit{More efficient CSIT-free schemes for massive WET:} EH deployments are usually static, and positioning information may be available. As such, novel and more efficient CSIT-free schemes can be designed to exploit such information. & Novel location-based designs relying on proper signals shifting over consecutive antennas, PB orientation and antenna array topology  are required for optimizing the EH processes. & 	environment geometry, location, QoS requirements & optimization, ray tracing, energy trading	\\	
		\bottomrule		
	\end{tabular}\label{table}
	\vspace{-4mm}
\end{table*}
	%
	
%	\bibliographystyle{IEEEtran}
%	\bibliography{IEEEabrv,references}
	%
	\vspace{-4mm}

%
%\section*{}
%

\end{document}